\documentclass[aps,prd,amsmath
,eqsecnum            
,preprintnumbers
,nofootinbib         
,twocolumn         
]{revtex4}
\usepackage[dvips]{graphicx}
\usepackage{amsmath}
\usepackage{amsfonts}
\usepackage{amssymb}
\usepackage{longtable}   

\allowdisplaybreaks[4]     

\begin{document}

\title{Redshift-drift in axially symmetric quasi-spherical Szekeres models}

\author{Priti Mishra}
\affiliation{Department of Astronomy and Astrophysics, \\
Tata Institute of Fundamental Research, \\
Homi Bhabha Road, Colaba, Mumbai -  400005, Maharashtra, India}
\email{priti@tifr.res.in}

\author{Marie-No\"elle C\'el\'erier}
\affiliation{Laboratoire Univers et Th\'eories (LUTH), \\
Observatoire de Paris, CNRS, Universit\'e Paris-Diderot \\
5 place Jules Janssen, 92190 Meudon, France}
\email{marie-noelle.celerier@obspm.fr}

\author{Tejinder P. Singh}
\affiliation{
Department of Astronomy and Astrophysics, \\
Tata Institute of Fundamental Research, \\
Homi Bhabha Road, Colaba, Mumbai -  400005, Maharashtra, India}
\email{tpsingh@tifr.res.in}

\date { }

\begin{abstract}
Models of inhomogeneous universes constructed with exact solutions of Einstein's General Relativity have been proposed in the 
literature with the aim of reproducing the cosmological data without any need for a dark energy component. Besides large scale 
inhomogeneity models spherically symmetric around the observer, Swiss-cheese models have also been studied. Among them, 
Swiss-cheeses where the inhomogeneous patches are modeled by different particular Szekeres solutions have been used for 
reproducing the apparent dimming of the type Ia supernovae (SNIa). However, the problem of fitting such models to the SNIa 
data is completely degenerate and we need other constraints to fully characterize them. One of the tests which is known to 
be able to discriminate between different cosmological models is the redshift-drift. This drift has already been calculated 
by different authors for Lema\^itre-Tolman-Bondi (LTB) models. We compute it here for one particular axially symmetric 
quasi-spherical Szekeres (QSS) Swiss-cheese which has previously been shown to reproduce to a good accuracy the SNIa data, 
and we compare the results to the drift in the $\Lambda$CDM model and in some LTB models that can be found in the literature. 
We show that it is a good discriminator between them. Then, we discuss our model's remaining degrees of freedom and propose a 
recipe to fully constrain them.
\end{abstract}

\maketitle


\section{Introduction} \label{sec1}

The last two decades have witnessed a phenomenal increase of the available cosmological data. Analyzed in the framework of an 
FLRW homogeneous cosmology, they have yielded the Concordance model where more than 95\% of the Universe content is of unknown 
nature. Among these 95\%, around 75\% of the Universe energy density is ascribed to the influence of some rather exotic component 
called dark energy. But dark energy has never been directly observed neither in the Universe, nor in laboratories, and it has very 
exotic properties, namely, it is a kind of fluid with negative pressure or a cosmological constant with an amplitude too small to
 account for the vacuum energy in the current standard model of particle physics. This has led some authors to studying whether 
the observations could not be given a more natural explanation in the framework of inhomogeneous models constructed with exact 
solutions of Einstein's field equations without any dark energy component.

The models most often found in the literature are roughly of two kinds: one patch large scale inhomogeneous Lema\^itre-Tolman-Bondi 
(LTB) models, spherically symmetric around the observer 
\cite{celerier00,iguchi02,alnes06,chung06,enqvist07,bolejko08a,garciabellido08,garciabellido08b,zibin08,yoo08,
enqvist08,ABNV09,BW09,CBK10,BNV10,MZS11,NaSa11}, and Swiss-cheese models in which the patches can be either 
spherically symmetric LTB holes \cite{KK07,MK07,BT07,MK08,BTT08,BT08,BN08} or non spherical Szekeres patches \cite{BC10,BS11}.

Since the LTB solutions are determined by two free functions of the radial coordinate plus a radial coordinate choice, 
two independent sets of data are necessary and sufficient to define them, e.g. angular distance and galaxy number counts 
\cite{SEN92,RM96,NHE97,AS99,CBK10}. In the most recent works devoted to solve the dark energy problem with  
zero-$\Lambda$ LTB models, such solutions have been constrained by two or more sets of data measured on our past light cone \cite{alnes06}, 
 \cite{garciabellido08, garciabellido08b, zibin08}, \cite{ABNV09, BW09}, \cite{BNV10,MZS11,NaSa11},
\cite{ZGR12,ZM11,BCF12} . By 
construction, we are thus left with a degeneracy as regards the $\Lambda$CDM model, because in both cases, homogeneous 
and inhomogeneous cosmology, the same data are reproduced without the possibility to discriminate between the models. 
However, it has been suggested that this degeneracy can only occur if the LTB model is not smooth at the center; 
otherwise, the models are 
distinguishable \cite{CFL08, AER11}. 

The degeneracy is even worse, when a few independent data sets are used, for models constructed with Szekeres solutions 
which are determined by five free functions of the radial coordinate plus a radial coordinate choice in the most general 
case. Even if this number can be reduced to three in the case of axial symmetry, as we discuss in Sec. \ref{sec5}, 
the mere supernova data used to constrain the models in \cite{BC10} is far from being sufficient for fully determining them.

This is the reason why tests using the redshift-drift have been recently proposed to deal with this issue. The redshift-drift 
is the temporal variation of the redshift of distant sources when the observation of the same source is done at 
observer's different proper times in an expanding universe. It allows one to make observations on the past light cones 
of an observer at different cosmological times, therefore giving access to a slice of space-time. It is thus a good 
discriminator between different cosmological models able to reproduce the observational data on our current past light 
cone. This effect has first been considered by Sandage \cite{Sandage62}, then by McVittie \cite{MV62}. In 
Friedmann-Lema\^itre-Robertson-Walker (FLRW) models, when the Universe expansion decelerates, all redshifts 
decrease with time. In FLRW models where the expansion is recently accelerating, like in the $\Lambda$CDM model, 
sources with redshifts $\lesssim 2.5$ exhibit a positive redshift-drift. In \cite{Uzan08} this effect has been proposed 
to test the ``Copernican Principle'' and to close the reconstruction problem of an LTB model from the luminosity distance 
as inferred from the supernova data. Other authors have since examined this effect for one patch zero-$\Lambda$ LTB models 
\cite{QA10,Yoo11}. It has been shown in \cite{MNC12} that this effect is the only currently proposed one able to test, 
in principle, the LTB one patch models against the $\Lambda$CDM model ``outside the past light cone'', provided spherical 
symmetry is but a mathematical simplification and one considers LTB models as exhibiting an energy density smoothed out over 
angles around us. However, other proposals designed to test specific LTB models considered  as a single exact spacetime and 
relying on conditions inside the observer's light cone can be found in the literature, e.g. tests using the BAO scale 
\cite{MZS11,ZGR12}, the kinematic Sunyaev-Zel'dovich effect \cite{garciabellido08b,ZM11,BCF12} or the Compton 
y-distortion of the CMB spectrum \cite{MZS11,JPZ11}.

Here, we want to study the redshift-drift in Swiss-cheese models which seem to represent more closely our observed Universe
 with its voids and filamentary patterns.

For voids, the spherically symmetric shape of the holes can be considered as a good approximation since non spherical voids 
evolve towards a spherical configuration. Actually, it has been shown, using the top-hat void model, that the 
smaller axis of an underdense ellipsoid stretches out faster than the longer one implying that voids become 
increasingly spherical as they evolve \cite{SW04}. Conversely, overdense ellipsoids tend to form pancakes and 
filamentary structures which are what we observe in the Universe. This is the reason why we choose to study a 
Swiss-cheese with non spherically symmetric overdense patches.

An inhomogeneous exact solution of General Relativity with no symmetry at all (no Killing vector) is the 
Szekeres model \cite{Sz75}. Among its sub-classes, the quasi-spherical Szekeres (QSS) case is best suited for 
our purpose since it possesses all three FLRW classes of models as an homogeneous limit and can therefore be 
matched to any kind of FLRW background. However, it has been shown in \cite{KB11,AK11} that in a general 
Szekeres model, generic light rays do not have repeatable paths (RLPs), i.e. two rays sent from the same 
source at different times to the same observer do not proceed through the same succession of intermediate
 matter particles, implying thus an angular drift of the source on the sky. This implies that, in such models, 
the light rays emitted from a source at different time coordinates are emitted in different directions and 
therefore reach different loci at the border of each patch. Since we integrate the null geodesic and redshift-drift equations for a sequence of such different time coordinates for each source, i.e. for each redshift $z$, 
this makes impossible the choice of the locus of the matching between two patches and thus impairs the 
construction of Swiss-cheese models. 
There are only two Szekeres classes,
 besides FLRW models, where RLPs exist, the most interesting for our purpose being the axially symmetric 
Szekeres models, in which the RLPs are the null geodesics intersecting every space of constant time on the axis of symmetry.

Actually, such a class of models have been used in \cite{BC10} (hereafter BC) to construct QSS Swiss-cheeses 
which can reproduce the supernova data. The model which we consider here is BC's model 5 which best reproduces 
these data and which is a model with overdense patches matched to a open FLRW background. In this model, the 
observer is located at the origin of the first patch and the sources, at redshifts ranging from $z=0$ 
to $z\sim 2.5$, can be anywhere on null geodesics axially directed towards the observer. Of course, this 
Swiss-cheese model must be still considered as a very simplified toy-model since axial symmetry is not a 
generic property of structures in the observed Universe.

Anyhow, it is important to compute the redshift-drift in such configurations and to compare it to the 
results in the $\Lambda$CDM model and other LTB models.

Inhomogeneous exact models have been sometimes criticized on the ground that they exhibit more degrees of freedom 
than FLRW models and are thus able to fit more easily the data.  Another interesting question to be addressed is 
therefore: how many independent data sets do we need to fully reconstruct an axially symmetric QSS model?

It has already been shown in \cite{Uzan08} that the combination of luminosity distance and redshift-drift data allows 
one to fully constrain a spherically symmetric spacetime. This applies, in particular, to LTB models. It has 
also been shown in \cite{NHE97} that, for any given isotropic observations of the apparent luminosity $l(z)$ 
and of the galaxy number count $n(z)$, with any given source evolution functions $\hat{L}(z)$ and $\hat{m}(z)$, 
a set of functions determining a zero-$\Lambda$ LTB model can be found to make the LTB observational relations fit the observations.

We demonstrate here, as a first step towards a more general theorem to be applied to QSS models without any symmetry, 
that an axially symmetric zero-$\Lambda$ Szekeres model needs three independent observation sets to be fully reconstructed.

The structure of the present paper is as follows. In Sec.~\ref{sec2}, we present the Szekeres models and the particular 
QSS subclass used in this paper. Section \ref{sec3} is devoted to the derivation of the differential equation giving 
the redshift-drift. In Sec.~\ref{sec4}, we display our result for the computation of the drift in BC model 5 and we 
compare it to that of the $\Lambda$CDM model and other LTB models studied in the literature. In Sec.~\ref{sec5}, we 
discuss the issue of closing the reconstruction of our type of model with a sufficient number of data sets, and we 
propose a solution for axially symmetric QSS models. In Sec.~\ref{sec6}, we present our conclusions.

\section{Szekeres models} \label{sec2}

The Szekeres solutions \cite{Sz75} are the most general solutions of Einstein's equations one can obtain with a dust 
gravitational source. They have no symmetry, i.e., no Killing vector, and are therefore well-suited to describe a 
lumpy universe. Their metric in comoving coordinates and synchronous time gauge is

\begin{equation}
{\rm d} s^2 =  c^2 {\rm d} t^2 - {\rm e}^{2 \alpha} {\rm d} r^2 - {\rm e}^{2 \beta} ({\rm d}x^2 + {\rm d}y^2),
\label{metsz}
\end{equation}
where $\alpha$ and $\beta$ are functions of $(t,r,x,y)$ to be determined by the field equations.

There are two families of Szekeres solutions. The class II family, where $\beta' = 0$ (here the prime denotes 
derivative with respect to $r$), is a simultaneous generalization of the Friedmann and Kantowski-Sachs models. 
Its spherically symmetric limit is the Datt-Ruban solution whose physical interpretation is not clear \cite{VR68,VR69}. 
The class I family of solutions 
is obtained when $\beta' \neq 0$. They contain the LTB solution as a spherically symmetric limit. 
We choose therefore this class of solutions to study the redshift-drift in Szekeres models. After solving the 
Einstein equations, its metric can be written, after a change of parametrization more convenient for our purpose \cite{H96},

\begin{equation}
{\rm d} s^2 =  c^2 {\rm d} t^2 - \frac{(\Phi' - \Phi {  E}'/ {  E})^2}
{\epsilon - k} {\rm d} r^2 - \Phi^2 \frac{({\rm d} x^2 + {\rm d} y^2)}{{  E}^2},
\label{metsz2}
 \end{equation}
where $\epsilon = 0, \pm 1$, $\Phi$ is a function of $t$ and $r$, $k$ is a function of $r$, and
\begin{equation}
{  E} = \frac{S}{2} \left[ \left( \frac{x-P}{S} \right)^2
+ \left( \frac{y-Q}{S} \right)^2 + \epsilon \right],
\label{Edef}
\end{equation}
with $S(r)$, $P(r)$, $Q(r)$, functions of $r$.

\subsection{Quasi-spherical Szekeres models}

As it appears from (\ref{metsz2}), only $\epsilon = +1$ allows the solution to have 
the three Friedmann limits (hyperbolic, flat and spherical). This is induced by the requirement of 
a Lorentzian signature for the metric. Since we are interested in the Friedmann limit of our model 
which we expect to become homogeneous at very large scales, i.e., that of the last-scattering, 
we focus only on the $\epsilon = +1$ case. It is called the quasi-spherical Szekeres (QSS) model.

Its metric, obtained with $\epsilon = +1$ in eq.(\ref{metsz2}), becomes
\begin{equation}
{\rm d} s^2 =  c^2 {\rm d} t^2 - \frac{(\Phi' - \Phi {  E}'/ {  E})^2}
{1 - k} {\rm d} r^2 - \Phi^2 \frac{({\rm d} x^2 + {\rm d} y^2)}{{  E}^2},
\label{ds2}
 \end{equation}
where
\begin{equation}
{  E} = \frac{S}{2} \left[ \left( \frac{x-P}{S} \right)^2
+ \left( \frac{y-Q}{S} \right)^2 +1 \right].
\end{equation}

Applying the Einstein equations to the metric (\ref{ds2}) and assuming the energy momentum tensor is that of dust, the Einstein equations reduce to the following two:

\begin{equation}
\frac{1}{c^2}\dot{\Phi}^2 = \frac{2M}{\Phi} - k + \frac{1}{3} \Lambda
\Phi^2, \label{vel}
\end{equation}
where the dot denotes derivation with respect to $t$, $\Lambda$ is the cosmological constant and $M(r)$ is an arbitrary function of $r$ related to the density $\rho$ via
\begin{equation}
\kappa \rho c^2= 
 \frac{2M' - 6 M {  E}'/{  E}}{\Phi^2 ( \Phi' - \Phi {  E}'/{  E})}, \label{rho}
\end{equation}
where $\kappa=8\pi G/c^4$.

The 3D Ricci scalar is
\begin{equation}
^{3}\mathcal{R} = 2 \frac{k}{\Phi^2} \left( \frac{ \Phi k'/k - 2 \Phi
{  E}'/{  E}}{ \Phi' - \Phi {  E}'/{  E}} + 1 \right).
\label{3dr}
\end{equation}
The Weyl curvature tensor decomposed into its electric and magnetic part is
\begin{eqnarray}\label{Weyl}
&& E^{\alpha}{}_{\beta} = C^{\alpha}{}_{\gamma \beta \delta} u^{\gamma} u^{\delta} =
\frac{M(3 \Phi' - \Phi M'/M)}{3 \Phi^3 ( \Phi' - \Phi E' / E)}
{\rm diag} (0,2,-1,-1), \nonumber \\
&& H_{\alpha \beta} = \frac{1}{2} \sqrt{-g} \epsilon_{\alpha \gamma \mu \nu} C^{\mu
\nu}{}_{\beta \delta} u^{\gamma} u^{\delta} = 0,
\end{eqnarray}
 where $\epsilon_{\alpha \gamma \mu \nu}$ is the 4-dimensional Levi-Civita symbol.
It can be easily noticed that (\ref{vel}) -- (\ref{Weyl}) reduces to the LTB equations once $E'/E$ is set to vanish.

Since (\ref{vel}) is the same in this model as in the LTB model, the bang time function, $t_B(r)$, follows from (\ref{vel}) in the same manner as
\begin{equation}
\int\limits_0^{\Phi}\frac{{\rm d} \widetilde{\Phi}}{\sqrt{- k + 2M /
\widetilde{\Phi} + \frac 1 3 \Lambda \widetilde{\Phi}^2}} = c [ t - t_B(r)].
\label{tbf}
\end{equation}

All the formulas given so far being covariant under coordinate transformations of the form $\tilde{r} = g(r)$, this means that one of the functions $k(r)$, $S(r)$, $P(r)$, $Q(r)$, $M(r)$ and $t_B(r)$ can be fixed at our convenience by the choice of $g$. Hence, each Szekeres solution is fully determined by only five functions of $r$. In the following, we choose $S$, $P$, $Q$, $M$ and $t_B$, and we make the coordinate choice: $\Phi(t_0, r) = r$.

\subsection{Axially symmetric QSS models. Null cone equations.}

We have seen in Sec.~\ref{sec1} that only axially symmetric QSS models possess RLPs and are therefore more suited for an easy study of the redshift-drift in Swiss-cheese models. Moreover, in these models, the only RLPs are axially directed null geodesics. Since, in our Swiss-cheese models, we use only radially directed light rays, this implies, as shown in \cite{ND07} (see also \cite{BKHC10}), that the Szekeres model should be axially symmetric and that, accordingly, the rays we consider for the computation of the drift are RLPs. The simplest axially-symmetric Szekeres model obeys

\[ P(r) = x_0 = {\rm ~const}, \quad \quad Q(r) = y_0 = {\rm ~const}. \]
In this case the dipole axis is along $x=x_0$ and $y=y_0$ (or in spherical coordinates along the directions $\vartheta=0$ and $\vartheta=-\pi$).

For the axially directed geodesics (${\rm d} x = {\rm d} y = 0$), we obtain from the null condition in (\ref{ds2})

\begin{equation}
 \frac{{\rm d} t}{{\rm d} r} = \pm \frac{1}{c} \frac{\Phi' - \Phi {  E}'/{  E}}{\sqrt{1 - k}}.
\label{snge}
\end{equation}
The plus sign is for $r_e < r_o$ and the minus sign for $r_e > r_o$, with $r_e$,
 the radial coordinate of the source and $r_o$, the radial coordinate of the observer. 
Since we put the observer at the origin $r_o=0$, we use the minus sign in (\ref{snge}).

The redshift relation in this case is \cite{BKHC10}

\begin{equation}
\ln (1+z) = - \frac{1}{c} \int\limits_{r_e}^{r_o} {\rm d} r
\frac{ \dot{\Phi}' - \dot{\Phi} {  E}'/{  E}}{\sqrt{1 - k}},
\label{srf}
\end{equation}
or equivalently,
\begin{eqnarray}
 \frac{{\rm d} r}{{\rm d} z} = \frac{c}{1+z} \frac{\sqrt{1-k}}{
\dot{\Phi}' - \dot{\Phi} {  E}'/{  E}}, \nonumber \\
 \frac{{\rm d} t}{{\rm d} z} = -  \frac{1}{1+z} \frac{\Phi' - \Phi {  E}'/{  E}}{
\dot{\Phi}' - \dot{\Phi} {  E}'/{  E}}.
\label{redrel}
\end{eqnarray}

\section{The redshift-drift equation for an axially symmetric QSS model} \label{sec3}

The redshift-drift is the redshift increase or decrease a comoving observer looking
 at the same comoving source on her past light cone can measure while her proper time elapses.
 That means that the redshift of the source is measured on the observer's two different past light cones.

\begin{figure}[!htb]
 \begin{center}
 \includegraphics[width=9cm]{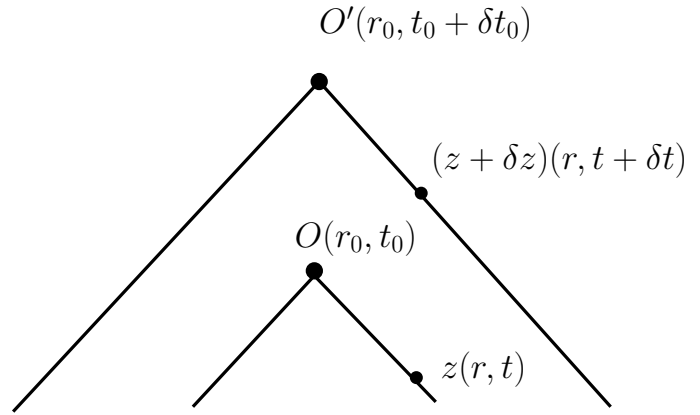}
\caption{The redshift-drift $\delta z$ of a source, initially at a redshift $z$ on the past
 light cone of an observer at $O$, as measured by the same observer at $O'$ after an elapsed 
time $\delta t_0$ of the observer's proper time.}
\end{center}
\end{figure}

An equation giving the redshift-drift in LTB models was derived in \cite{Yoo11}. We 
adapt here the reasoning to obtain such an equation for the axially symmetric QSS model.

We consider an observer $O$ located at the origin, with coordinates $(t_0, r=0)$. After 
an elapsed time $\delta t_0$, this comoving observer is at $O'$ $(t_0 + \delta t_0, r=0)$. 
A comoving source, with radial coordinate $r$ has, on the past light cone issued from $O$, 
redshift $z(r)$ and time coordinate $t(r)$. The same comoving source has, on the light cone issued from $O'$, redshift
\begin{equation}
Z(r) = z(r) + \delta z(r),
\label{eq1}
\end{equation}
and time coordinate
\begin{equation}
T(r) = t(r) + \delta t(r),
\label{eq2}
\end{equation}
with $t(r=0) = t_0$, $z(r=0) = Z(r=0) = 0$, $\delta z(r=0) = 0$ and $\delta t(r=0) = \delta t_0$.

From the geodesic equations (\ref{redrel}), the equation for the redshift is
\begin{equation}
\frac{{\rm d}z}{{\rm d}r} = \frac{1 + z}{c} \frac{\dot{\Phi}' - \dot{\Phi}E'/E}{\sqrt{1 - k}}.
\label{eq3}
\end{equation}

Eq.~(\ref{eq1}) can be written
\begin{equation}
\frac{{\rm d}\delta z(r)}{{\rm d}r} = \frac{{\rm d}Z(r)}{{\rm d}r} - \frac{{\rm d}z(r)}{{\rm d}r}.
\label{eq5}
\end{equation}

Inserting (\ref{eq3}) in (\ref{eq5}) and considering $\delta z(r)$ and $\delta t(r)$ as negligible with respect to $z$ and $t$, we keep only the first order terms in $\delta z$ and $\delta t$ and obtain
\begin{eqnarray}
\frac{{\rm d}\delta z}{{\rm d}r} &=& \frac{1 + z}{c} \frac{\left[\ddot{\Phi}'(t,r) - \ddot{\Phi}(t,r)E'/E\right]}{\sqrt{1 - k}} \delta t \\ \nonumber
&+& \frac{\dot{\Phi}'(t,r) - \dot{\Phi}(t,r)E'/E}{\sqrt{1 - k}} \frac{\delta z}{c}.
\label{eq8}
\end{eqnarray}

Now, (\ref{eq2}) can be written
\begin{equation}
\frac{{\rm d}\delta t(r)}{{\rm d}r} = \frac{{\rm d}T(r)}{{\rm d}r} - \frac{{\rm d}t(r)}{{\rm d}r}.
\label{eq9}
\end{equation}
Inserting (\ref{snge}) with the minus sign in (\ref{eq9}) and considering $\delta t(r)$
 as negligible with respect to $t$, therefore keeping only the first order term in $\delta t$, we obtain
\begin{equation}
\frac{{\rm d}\delta t(r)}{{\rm d}r} = - \frac{1}{c} \frac{\dot{\Phi}'(t,r) - \dot{\Phi}(t,r)E'/E}{\sqrt{1 - k}} \, \delta t
\label{eq11}
\end{equation}

We consider the case where the redshift $z$ is monotonically increasing with $r$ 
and we replace the independent variable $r$ by $z$. By using
\begin{equation}
\frac{{\rm d}}{{\rm d}r} = \frac{{\rm d}z}{{\rm d}r} \frac{{\rm d}}{{\rm d}z} = 
\frac{1 + z}{c} \frac{\dot{\Phi}' - \dot{\Phi}E'/E}{\sqrt{1 - k}} \frac{{\rm d}}{{\rm d}z}
\label{eq12}
\end{equation}
in (\ref{eq8}) and after some rearrangement, we obtain
\begin{equation}
\frac{{\rm d}\delta z}{{\rm d}z} = \frac{\ddot{\Phi}' - \ddot{\Phi}E'/E}{\dot{\Phi}' - 
\dot{\Phi}E'/E} \delta t + \frac{\delta z}{1 + z}.
\label{eq14}
\end{equation}

Now, we insert (\ref{eq12}) into (\ref{eq11}), that gives
\begin{equation}
\frac{{\rm d}\delta t}{{\rm d}z} = - \frac{\delta t}{1 + z},
\label{eq16}
\end{equation}
which we integrate from the observer O at $(t_0, z=0)$ to the source at $(t, z)$ to obtain
\begin{equation}
\delta t = \frac{\delta t_0}{1 + z}.
\label{eq17}
\end{equation}

We insert this expression for $\delta t$ into (\ref{eq14}) and obtain the equation for the redshift-drift,
\begin{equation}
\frac{{\rm d}}{{\rm d}z} \left(\frac{\delta z}{1 + z}\right) = \frac{1}{(1 + z)^2} \frac{\ddot{\Phi}' - 
\ddot{\Phi}E'/E}{\dot{\Phi}' - \dot{\Phi}E'/E} \delta t_0.
\label{eq19}
\end{equation}
Once $\delta z$ is obtained by numerically integrating (\ref{eq19}) for a fixed value of $\delta t_0$, 
the redshift-drift follows from its definition $\dot{z}=\delta z / \delta t_0$.

\section{Computation of the redshift-drift in BC's Swiss-cheese model 5} \label{sec4}

\subsection{The Swiss-cheese model}

As seen above, in the case of axially directed geodesics the equations which describe light 
propagation and redshift-drift simplify significantly. Moreover, density fluctuations (\ref{rho}) 
and curvature fluctuations, (\ref{3dr}) and (\ref{Weyl}), are the largest along the axial axis. 
When a light ray passes through the origin
${  E}'/{  E} \rightarrow -{  E}'/{  E}$ \cite{PK06,BKHC10}. Since, in the above equations, $E'/E$ 
is always multiplied by $\Phi$ or $\dot{\Phi}$ which are zero at the
origin, there is no discontinuity here. Now, 
a much smaller discontinuity in ${ E}'/{ E}$ (which probably appears due to the imperfect matching in the BC Swiss Cheese model, 
and would probably not occur for a model with perfect matching), of order $4\times10^{-6}$ Mpc$^{-1}$ 
(to be compared with the larger 
discontinuity of order $10^{-3}$ Mpc$^{-1}$ at the origin) occurs also 
at the boundaries. However, its magnitude is not sufficient to impact noticeably the results for 
the redshift-drift as can be seen in Fig.~\ref{drift_compfig}.

When constructing a Swiss-cheese model, one needs to satisfy the junction conditions at the patch borders. 
In our models the Szekeres inhomogeneities (the patches) are matched to the Friedmann background (the cheese). 
These Szekeres patches are arranged so that their boundaries touch wherever a light ray exits its neighbor. 
Thus the ray immediately enters another patch and spends no time in the Friedmann background.
The matching conditions across a comoving spherical surface, $r =$ constant, are: the mass inside the junction 
surface in the Szekeres patch is equal to the mass that would be inside that surface in the homogeneous 
background (this mass is defined as the mass energy density $\rho$ integrated over the spatial volume inside 
the junction surface at a fixed time coordinate t); the spatial curvature at the junction surface is 
the same in both the Szekeres and Friedmann 
regions, which implies $k_{SZ} = k_F r^2$ and $(k_{SZ})' = 2 k_F r$; the bang time and $\Lambda$ must be 
continuous across the junction. In our model, $\Lambda=0$ in both the patches and the cheese.

Besides matching the inhomogeneous patches, one also needs to take care of the null geodesics. However, 
since we only consider here axial geodesics, the junction is trivial and requires only matching the radial,
 or equivalently, the time component \cite{KB09}.

The Swiss-cheese model we consider exhibits patches that are described by the following functions:

\begin{equation}
M = \frac{\rho_b \kappa c^2}{2} \int\limits_0^r {\rm d} \tilde{r} \, \tilde{r}^2  \left[ 1 + \delta_{\rho} - \delta_{\rho} \exp \left( -
\frac{\tilde{r}^2}{\sigma^2} \right) \right],
\end{equation}
where 
$\rho_{b} =  \Omega_m \frac{3H_0^2}{8 \pi G}$,
$\Omega_m = 0.3$, $H_0=68$ km s$^{-1}$ Mpc$^{-1}$,
$\delta_{\rho} = 0.6$ and $\sigma = 180$ Mpc, $t_B=P=Q=0$ and
$S= (10^3 + \ell)^{\pm 0.8}$, with $\ell = r/$kpc and where $-$ is for propagation from the origin with 
${  E}'/{  E} = S'/S = - 0.8 /(10^3 + \ell)$ and $+$ towards the origin with ${  E}'/{  E} = S'/S = 0.8 /(10^3 + \ell)$. 

As can be shown from (\ref{rho})--(\ref{Weyl}), for $r>400$ Mpc this model becomes almost homogeneous. 
We join inhomogeneous patches at $r=400$Mpc. By construction, this is not a perfect matching. Only the 
functions $M$ and $k$ are continuous along 
this boundary but not for example the curvature [eqs. (\ref{3dr}) and (\ref{Weyl})]. In BC's paper, this non 
perfect matching is investigated with the use of a so-called minimum void scenario which exhibits the same 
arbitrary functions as model 5 but with only one inhomogeneous patch, and therefore no matching. The result 
is that the $\chi^2$s for reproducing the SNIa data are exactly the same in both models, the Swiss-cheese model 5 
and minimal void model 5, and are equal to 269 for 307 measurements. This shows that the non-perfect matching has no impact at all 
on SNIa data fitting. Our result for the redshift-drift, displayed in Fig.~\ref{drift_compfig}, shows that this non-perfect 
matching does not impair significantly the calculation of this effect (no visible discontinuities 
of the derivative at the boundaries there). This ensures the validity of our model for the computation of the redshift-drift.

We place the observer at the origin $r=0$ of the first patch and we consider the light emitted by the 
sources in the first patch and directed towards the observer. As we have seen above, in this case 
${ E}'/{ E}$ is positive. After calculating the redshift-drift for all the comoving sources with 
$0 < r < 400$ Mpc, we are led to the second patch and consider the sources located between the
 border of this patch and its own origin. The light traveling towards the observer moves away 
from this origin and we have therefore to change the sign of ${ E}'/{ E}$. The same occurs when 
we are led to the other side of the second patch origin. Light emitted from there travels again 
towards an origin and the sign of ${ E}'/{ E}$ needs once again to be inverted, and so on.

\subsection{The algorithm}

In order to calculate the redshift-drift, we proceed in the following manner:
\begin{enumerate}
\item To find $k(r)$ we proceed as follows:

After substituting $t=t_0$ and $t_B(r)=0$ into (\ref{tbf}) we fix a value of $r$, which is 
the upper limit of the integral, between $0$ and $400$ Mpc.
We want such a $k$ which satisfies (\ref{tbf}).
We find that to have a real value of the integrand over the whole range of integration $k<2M/r$ is necessary.
But in the range $0<k<2M/r$, we do not obtain any root of (\ref{tbf}). Therefore we choose the case $k(r)<0$ 
for which the parametric solution is the same as in the LTB model and is given by  
\begin{equation} 
\Phi(r,t)=\frac{M}{(-k)}(\cosh\eta-1)
\label{solutionforeeq1}
\end{equation}
and 
\begin{equation} 
t-t_B(r)=\frac{M}{(-k)^{3/2}}(\sinh\eta-\eta)
\label{solutionforeeq2}
\end{equation}
where $\eta(t,r)$ is the parameter. 

\item Using the initial condition $\Phi(t_0,r) = r$ in (\ref{solutionforeeq1}) and 
substituting $\eta=\eta_0$ at $t=t_0$, we get $k(r)$ as
\begin{equation} 
-k(r)=\frac{M(r)(\cosh\eta_0-1)}{r}
\end{equation}

\item We obtain  $\eta_0$ by solving numerically (\ref{solutionforeeq2}) for $t=t_0$. 
The numerical value of $t_0$ for our model is 11.63 Gyrs.

\item Then we find $t(r)$ on the past light cone by numerically solving the null condition equation 
(\ref{snge}) for incoming geodesics. The sign of $E'/E$ is chosen as positive(negative) depending on 
whether a light ray is moving towards(away from) the origin in the patch in which it is.

\item 
Substituting $t(r)$ in (\ref{solutionforeeq2}) we obtain $\eta(r)$ on the past light cone issued from $O$, 
using which in (\ref{solutionforeeq1}) we calculate $\Phi(t(r),r)$ and its derivatives on the past light cone.

\item Then we numerically solve the first of the two equations (\ref{redrel}) for the redshift $z(t(r),r)$.

\item After having found $z$, we find the redshift-drift at this $z$ by numerically solving (\ref{eq19}).

\end{enumerate}

The various functions obtained by this procedure are described in the next sub-section.

\subsection{The results}

\begin{figure}[!htb]
\begin{center}
 \includegraphics[width=9cm]{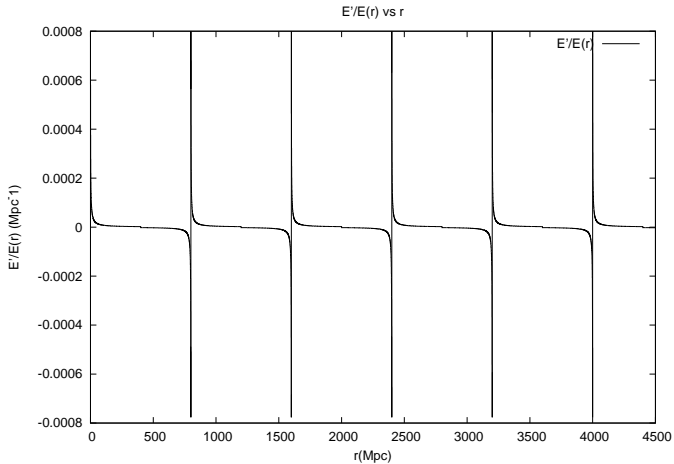}
\caption{The $E'/E(r)$ function as a function of the radial coordinate $r$.}
\label{eprimeoverfig}
\end{center}
\end{figure}

\begin{figure}[!htb]
\begin{center} 
 \includegraphics[width=9cm]{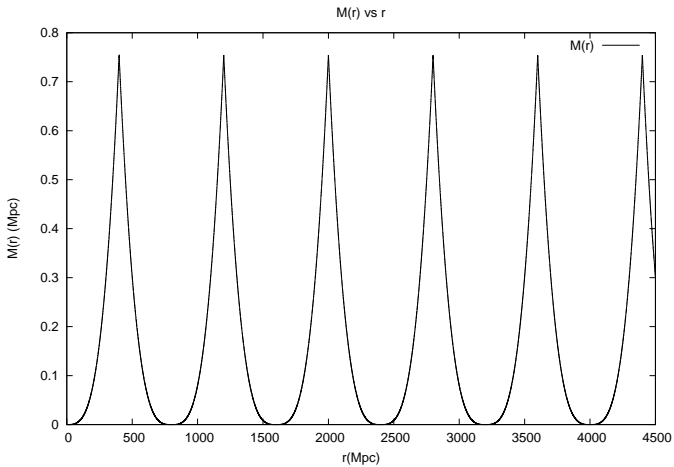}
\caption{The effective gravitational mass $M(r)$ as a function of the radial coordinate $r$.}
\label{Mvsr}
\end{center}
\end{figure}

\begin{figure}[!htb]
\begin{center}
 \includegraphics[width=9cm]{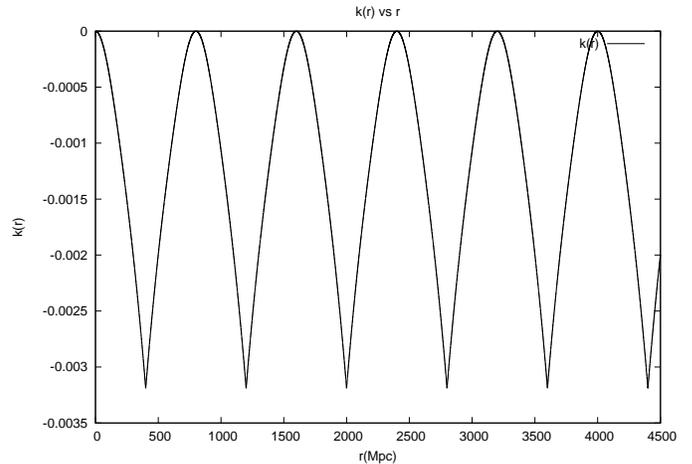}
\caption{The $k(r)$ function, determining the model evolution type, as a function of the radial coordinate $r$.}
\label{kvsr}
\end{center}
\end{figure}

\begin{figure}[!htb]
\begin{center}
 \includegraphics[width=9cm]{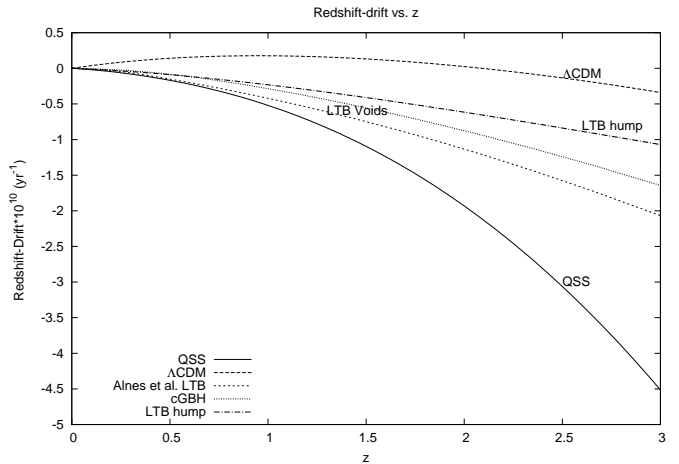}
\caption{The redshift-drift ($\delta z / \delta t_0$) as a function of the redshift $z$ for the axially symmetric QSS 
Swiss-cheese model, the $\Lambda$CDM model, the cGBH LTB void model  \cite{garciabellido08} (courtesy \cite{QA10}), 
the Alnes et al. LTB void model \cite{alnes06} (courtesy : \cite{QA10}) and the Yoo LTB hump model \cite{Yoo10} (courtesy : \cite{Yoo11})}
\label{drift_compfig}
\end{center}
\end{figure}

The results obtained by the numerical computation described above are plotted in Figs.~\ref{eprimeoverfig}--\ref{drift_compfig}. 
The radial coordinate extends up to about 4,500 Mpc (some 5.6 patches) and the redshift, up to 3.

Figure~\ref{eprimeoverfig} depicts the function $E'/E$ over these patches, while the functions $M(r)$ and $k(r)$
 are presented in Figs.~\ref{Mvsr} and \ref{kvsr} respectively.  Figure~\ref{drift_compfig} depicts the redshift-drift 
for the Szekeres model and compares it with that for the $\Lambda$CDM model, and for three LTB models: Alnes et al.'s void model \cite{alnes06} and the so-called constrained GBH (cGBH) void model \cite{garciabellido08}, both studied in \cite{QA10}, and Yoo's hump model \cite{Yoo10}, studied in \cite{Yoo11}. As is well-known, the drift 
for the $\Lambda$CDM model is positive up to some redshift \cite{Uzan08,QA10} - this is of course a distinguishing 
feature for an accelerating FLRW cosmology. 

The redshift-drift for our Szekeres Swiss-cheese model is negative while for 
the $\Lambda$CDM model it is positive until a redshift $~2.5$.
However, we wish to stress here that the minus sign of the drift appearing in our particular model cannot be 
considered as a general feature of Szekeres cosmology. This should actually be studied by an analysis extended 
to a number of other QSS Swiss-cheese models. This will be the subject of future work.

As stressed before, the redshift-drift in our Szekeres model is also compared, in Fig.~\ref{drift_compfig}, 
with that in three LTB models: two LTB void models, the Alnes et al.-type \cite{alnes06} model I and the cGBH 
model \cite{garciabellido08} studied in \cite{QA10}, and a LTB hump model described in \cite{Yoo10} whose 
redshift drift is shown in \cite{Yoo11}. In the void models, an inner under-density makes a smooth  transition 
to an outer region with higher density. In the hump model the density profile exhibits a large over-density near 
the center. In all three models the observer is centrally located. These models are constructed in order to 
reproduce to a very good accuracy different cosmological data sets among which the SNIa data. They show the 
same general shape for the drift curve as our Szekeres model does, with the drift becoming more negative with 
increasing $z$. Note however that the magnitude of the drift in our Szekeres model is higher by a factor of 
about two, at a given redshift, in comparison to the LTB models (at those redshifts where the LTB curves show 
a decline with increasing $z$). We want also to stress here that, even though the redshift drift of an off-center 
source observed at the symmetry center of any LTB void model is always negative, this is not obligatorily the case 
in a LTB hump model \cite{Yoo11}. The feature exhibited by the redshift drift in the particular hump model shown in 
Fig.~\ref{drift_compfig} cannot be therefore generalized to the whole class of such models.

\section{Fully constraining the model degrees of freedom} \label{sec5}

Inhomogeneous exact models have been sometimes criticized on the ground that they 
exhibit more degrees of freedom than FLRW models and are thus able to fit more easily 
the data. In Sec.~\ref{sec4}, we have studied a particular axially symmetric QSS 
Swiss-cheese model able to reproduce the supernova data and we have computed the 
redshift-drift for this model. However, another interesting question to be addressed 
is: how many and which independent data sets do we need to be able to fully reconstruct 
an axially symmetric Swiss-cheese QSS model from background observations?

Following \cite{ND07}, it has been shown in \cite{BKHC10} (Theorem 3.1) that a constant-$(x,y)$ null geodesic
 exists only in an axially symmetric Szekeres spacetime\footnote{We wish to stress here a mis-statement in 
this Theorem 3.1. It is claimed that there is only one null geodesic with constant $x$ and $y$ in the axially 
symmetric case. This is not true: the null geodesics that pass through the symmetry axis may have any radial 
direction, and may cross the symmetry axis at different instants of the cosmic time t. So, this is actually a 
2-parameter family of null geodesics, one parameter defining the azimuthal direction and the other defining 
the instant of intersection with the axis.}.

One can always find a coordinate system for $x$ and $y$ where, in such a spacetime, the functions $P$ and $Q$ 
obey the set of equations
\begin{equation}
(x_0, y_0) = (P, Q) \Longrightarrow P' = Q' = 0.
\label{pq}
\end{equation}

Hence, in these coordinates, the model is fully determined, not by six arbitrary functions of $r$, as in the 
most general Szekeres model, but only by  four. We choose here $M(r)$, $k(r)$, $t_B(r)$ and $S(r)$ or equivalently $E'/E(r)$.

Up to here, we have only at our disposal, to constrain our model, the supernova data used in BC and the redshift-drift 
we have computed in Sec.~\ref{sec3} to be used for such a purpose in the future. The question is: how many
 independent data sets do we need to fully constrain the four arbitrary functions of $r$ of such a model assuming 
these data are measured on the axial null geodesics directed towards the observer?

It has already been shown in \cite{Uzan08} that the combination of luminosity distance and redshift-drift data 
allows one to fully reconstruct a spherically symmetric spacetime. This applies, in particular, to LTB models. 
It has also been shown in \cite{NHE97} that, for any given isotropic observations of the apparent luminosity 
$l(z)$ and of the galaxy number count $n(z)$, with any given source evolution functions $\hat{L}(z)$ and 
$\hat{m}(z)$, a set of functions determining a zero-$\Lambda$ LTB model can be found to make the LTB 
observational relations fit the observations. We can reword this last statement as: the two arbitrary 
functions of $r$ of a zero-$\Lambda$ LTB model can be fully determined by two independent sets of observations 
realized on the observer past light cone.

A natural guess is therefore that an axially symmetric zero-$\Lambda$ Swiss-cheese Szekeres model needs three 
independent observation sets to be fully reconstructed. We demonstrate here, using a reasoning inspired from 
\cite{NHE97} \footnote{We wish to correct here an error in eq.(31) of \cite{NHE97}. At variance with what is 
claimed there, the luminosity distance is not the same as the diameter distance but is related to it by the 
reciprocity theorem \cite{IMHE33} expressed by our eq. (\ref{rec}).}, that this is indeed the case.

Note that, in this section, we work with units in which $c=G=1$.

\subsection{Coordinate choice}

We denote the past null cone issued from the observer at $(t=t_0,r=0)$ 
and composed of all the axially directed null geodesics of the Swiss-cheese 
by $t=\widehat{t}(r)$. From (\ref{snge}), $t=\widehat{t}$ satisfies,
\begin{equation}
\frac{{\rm d} \widehat{t}}{{\rm d} r} = - \frac{\Phi'[\widehat{t}(r),r] - 
\Phi[\widehat{t}(r),r] {E}'/{E}}{\sqrt{1 - k}} = - \frac{\widehat{\Phi'} - \widehat{\Phi} {E}'/{E}}{\sqrt{1 - k}},
\label{that}
\end{equation}
a hat denoting a quantity evaluated on the observer's past light cone.

We now use our freedom to choose the radial coordinate such that, on this past light cone,
\begin{equation}
\frac{\widehat{\Phi'} - \widehat{\Phi} {E}'/{E}}{\sqrt{1 - k}} = 1.
\label{coord}
\end{equation}
The equation for the incoming axial null geodesic is then
\begin{equation}
\widehat{t}(r) = t_0 -r
\label{ngeo}
\end{equation}

With this coordinate choice, (\ref{rho}) becomes
\begin{equation}
4 \pi \widehat{\rho} \widehat{\Phi}^2 = \frac{M' - 3M E'/E}{\sqrt{1 - k}},
\label{hatrho}
\end{equation}
and, on the past light cone, (\ref{vel}) must be written, with $\Lambda = 0$,
\begin{equation}
\widehat{\dot{\Phi}}^2 = \frac{2M}{\widehat{\Phi}} - k.
\label{hatdot}
\end{equation}
This equation possesses solutions in terms of a parameter ${\eta}(t,r)$ depending on the sign of $k$.
 When written on the past light cone, they become
\begin{equation}
\widehat{\Phi} = \frac{M(r)}{K(r)}\widehat{\phi}(t,r), \qquad \widehat{\xi}(t,r) = \frac{ K(r)^{3/2}[\widehat{t} - t_B(r)]} {M(r)}
\label{par}
\end{equation}

\begin{itemize}
 
\item For $k>0$,
\begin{equation}
K(r) = k \qquad \widehat{\phi} = 1 - \cos \widehat{\eta} \qquad \widehat{\xi} = \widehat{\eta} - \sin \widehat{\eta};
\label{para}
\end{equation}

\item For $k = 0$,
\begin{equation}
K(r) = 1 \qquad \widehat{\phi} = (1/2) \widehat{\eta}^2 \qquad \widehat{\xi} = (1/6) \widehat{\eta}^3;
\label{parb}
\end{equation}

\item For $k<0$,
\begin{equation}
K(r) = - k \qquad \widehat{\phi} = \cosh \widehat{\eta} - 1 \qquad \widehat{\xi} = \sinh \widehat{\eta} - \widehat{\eta}.
\label{parc}
\end{equation}

\end{itemize}

These expressions tell us the type of evolution for each region, based on the measured data: for $k>0$, the model expands away from an initial singularity and then re-collapses to a final singularity (elliptic evolution); with $k<0$ the model is either ever-expanding or ever-collapsing, depending on the initial conditions (hyperbolic evolution); $k=0$ is the intermediate case (parabolic evolution).

\subsection{Differential equations for the arbitrary functions} \label{diffeq}

The total derivative of $\Phi$ on the past null cone is
\begin{equation}
\frac{{\rm d} \widehat{\Phi}}{{\rm d} r} = \widehat{\Phi'} + \widehat{\dot{\Phi}} \frac{{\rm d} \widehat{t}}{{\rm d} r},
\label{totder}
\end{equation}
into which we substitute (\ref{coord}) and (\ref{ngeo}) to obtain
\begin{equation}
E'/E = \frac{1}{\widehat{\Phi}} \left[\frac{{\rm d} \widehat{\Phi}}{{\rm d} r} + \widehat{\dot{\Phi}} - \sqrt{1 - k}\right].
\label{derphi}
\end{equation}
From (\ref{hatdot}), we have
\begin{equation}
\widehat{\dot{\Phi}} = \pm \sqrt{\frac{2M}{\widehat{\Phi}} - k}.
\label{dotphi2}
\end{equation}
Since we are in an expanding Universe, we choose the + sign for $\widehat{\dot{\Phi}}$ in (\ref{dotphi2}). Thus, we substitute (\ref{dotphi2}) with the + sign into (\ref{derphi}) and obtain
\begin{equation}
E'/E = \frac{1}{\widehat{\Phi}} \left[\frac{{\rm d} \widehat{\Phi}}{{\rm d} r} + \sqrt{\frac{2M}{\widehat{\Phi}} -k} - \sqrt{1 - k}\right].
\label{E}
\end{equation}
This gives us a first order differential equation for the arbitrary function $E(r)$ which can be very easily integrated once $\widehat{\Phi}$ and the other arbitrary functions $M(r)$ and $k(r)$ are determined.

We wish to write now a first order differential equation for $M(r)$. We substitute (\ref{E}) into (\ref{hatrho}), which gives
\begin{eqnarray}
&&\frac{{\rm d}M}{{\rm d} r} - \frac{3M}{\widehat{\Phi}} \left[\frac{{\rm d} \widehat{\Phi}}{{\rm d} r} + \sqrt{\frac{2M}{\widehat{\Phi}} -k} - \sqrt{1 - k}\right] \nonumber \\
&=& 4 \pi \widehat{\rho} \widehat{\Phi}^2 \sqrt{1 - k}.
\label{diffM}
\end{eqnarray}
Note that, at variance with the differential equation for $M$ obtained in the LTB case in \cite{NHE97}, (\ref{diffM}) involves not only the observable quantity $\widehat{\rho}$, but also the unknown arbitrary function $k$. This will not invalidate our reasoning but make it a little more complicated.

We aim now at deriving a differential equation for $k$ involving observable quantities, the arbitrary function $M$, $\widehat{\Phi}$ and its total derivatives with respect to $r$.

The luminosity distance, $D_L$, of a source satisfies, on the past light cone of the observer,
\begin{equation}
\widehat{D_L}(z) = \sqrt{\frac{\widehat{L}(z)}{4 \pi \ell(z)}},
\label{dl}
\end{equation}
with $\widehat{L}$ being the absolute luminosity, i.e. the luminosity in the rest frame of the source, and $\ell(z)$, the measured 
bolometric flux, i.e. integrated over all frequencies, emitted by the source at redshift $z$.

The area distance, $D_A$, also known as the angular diameter distance, is related to the luminosity distance by the reciprocity theorem 
\cite{IMHE33},
\begin{equation}
D_A(z) = \frac{D_L(z)}{(1 + z)^2}.
\label{rec}
\end{equation}
Therefore, once we obtain a differential equation involving $\widehat{D_A}$, it is easy to transform it, through (\ref{rec}), into a
 differential equation involving $\widehat{D_L}$, which is a quantity measured on the observer's past light cone, as shown by (\ref{dl}).

It has been demonstrated in BC that, for axial geodesics, $D_A$ can be written
\begin{equation}
\frac{{\rm d^2} D_A}{{\rm d} s^2}  = - \frac{1}{2} R_{\alpha \beta} k^{\alpha} k^{\beta} D_A,
\label{darel}
\end{equation}
where $k^{\alpha}$ is the null tangent vector, $dx^{\alpha}/ds$, and $s$ is the  null affine parameter.
On an axially directed null geodesic $k^x = k^y = 0$ and (\ref{darel}) can thus be written
\begin{equation}
\frac{{\rm d^2} D_A}{{\rm d} s^2}  = - \frac{1}{2} \left[R_{tt}(k^t)^2 + 2 R_{tr}k^t k^r + R_{rr}(k^r)^2\right] D_A.
\label{das}
\end{equation}
We use
\begin{equation}
\frac{{\rm d^2} D_A}{{\rm d} s^2}  = \frac{{\rm d^2} D_A}{{\rm d} r^2} \left(k^r \right)^2 + \frac{{\rm d} D_A}{{\rm d} r} \frac{{\rm d} k^r}{{\rm d} s}.
\label{dda}
\end{equation}
With $E,_x = E,_y = E,_{rx} = E,_{ry} = 0$ on the axially directed null geodesics, the geodesic equation for $r$ becomes \cite{BKHC10}
\begin{equation}
\frac{{\rm d} k^r}{{\rm d} s} + 2 \frac{\dot{\Phi}_1}{\Phi_1} k^t k^r + \left[\frac{\Phi_1'}{\Phi_1} 
+ \frac{k'}{2(1-k)} \right] \left(k^r \right)^2 = 0,
\label{geor}
\end{equation} 

where
\begin{equation}
\Phi_1 \equiv \Phi' - \Phi {E}'/{E}.
\end{equation} 

With our coordinate choice, on the light cone, $k^r = -k^t$. Using (\ref{geor}), we insert (\ref{dda}) into (\ref{das}) and obtain
\begin{eqnarray}
 \frac{{\rm d^2} \widehat{D_A}}{{\rm d} r^2} &+& \left[ 2 \frac{(\widehat{\dot{\Phi}}_1)}{\widehat{\Phi}_1}
- \frac{\widehat{\Phi}_1' + \widehat{\Phi} (E'/E)^2}{\widehat{\Phi}_1} - \frac{k'}{2(1-k)} \right] \frac{{\rm d} \widehat{D_A}}{{\rm d} r} \nonumber \\
&+& \frac{1}{2} \left(\widehat{R_{tt}}  - 2\widehat{R_{tr}} + \widehat{R_{rr}}\right) \widehat{D_A} = 0.
\label{dar}
\end{eqnarray}

The $R_{tr}$ Ricci tensor component vanishes here and the $R_{tt}$ component is
\begin{equation}
R_{tt} = \frac{3 \ddot{\Phi}E'/E - \ddot{\Phi}' - 2 \ddot{\Phi} {\Phi}'/\Phi}{{\Phi}' - \Phi E'/E},
\label{rtt}
\end{equation}
and the $R_{rr}$ component can be written
\begin{eqnarray}
&& R_{rr} = \left(\frac{\Phi_1}{1-k}\right) \left(\ddot{\Phi}' - \ddot{\Phi} E'/E + 2 \dot{\Phi} \dot{\Phi}'/\Phi\right) \nonumber \\
&+& \frac{1}{1-k} \left\{ 2\left(\dot{\Phi}^2 -1 + k \right)(E'/E)^2 +2\frac{E'/E}{\Phi} \bigg[ - \Phi'\dot{\Phi}^2 \right. \nonumber \\
&+& \left. \Phi'(1-k) -k' \frac{\Phi}{2} \bigg] + k' \frac{\Phi'}{\Phi} \right\}.
\label{rrr}
\end{eqnarray}
Substituting (\ref{rtt}) and (\ref{rrr}) written on the light cone into (\ref{dar}), 
we obtain a differential equation involving $\widehat{D_A}(r)$, ${\rm d} \widehat{D_A}/{\rm d} r$ 
and ${\rm d^2} \widehat{D_A}/{\rm d} r^2$. However, what the observer measures on her past light 
cone are not these quantities, but rather $\widehat{D_L}(z)$ and its derivatives with respect to the redshift $z$. We apply therefore
\begin{equation}
\frac{{\rm d} \widehat{D_A}}{{\rm d} r} = \frac{{\rm d} \widehat{D_A}}{{\rm d} z} \frac{{\rm d} \widehat{z}}{{\rm d} r},
\label{partda}
\end{equation}
and
\begin{equation}
\frac{{\rm d^2} \widehat{D_A}}{{\rm d} r^2} = \frac{{\rm d} \widehat{D_A}}{{\rm d} z} \frac{{\rm d^2} 
\widehat{z}}{{\rm d} r^2} + \frac{{\rm d^2} \widehat{D_A}}{{\rm d} z^2} \left(\frac{{\rm d} \widehat{z}}{{\rm d} r}\right)^2,
\label{ppartda}
\end{equation}
while also using (\ref{eq3}), written on the light cone as
\begin{equation}
\frac{{\rm d}\widehat{z}}{{\rm d}r} = (1 + \widehat{z}) \frac{\widehat{\dot{\Phi}}_1}{\sqrt{1 - k}},
\label{red}
\end{equation}
and the $\widehat{D_L}$ expression as given by (\ref{rec}). We obtain, after some calculations
\begin{eqnarray}
&& \left(\frac{{\rm d} \widehat{D_L}} {{\rm d} z} - \frac{2 \widehat{D_L}} {1+\widehat{z}} \right) 
\bigg\{ \widehat{\dot{\Phi}}_1 \left[2\frac{\widehat{\dot{\Phi}}_1} 
{\widehat{\Phi}_1} -\frac{  \widehat{\Phi_1'}} {\widehat{\Phi}_1}+\frac{\widehat{\dot{\Phi}}_1} {\sqrt{1-k}}\right.  \nonumber \\
&+& \left. 
  \frac{1}{1+\widehat{z}} \right]  
+  \widehat{\dot{\Phi}_1'}  \bigg\}
+ \left[(1+\widehat{z})\frac{{\rm d^2} \widehat{D_L}} {{\rm d} z^2} - 4 \frac{{\rm d} \widehat{D_L}} {{\rm d} z} + \frac{6 \widehat{D_L}} {1+\widehat{z}} \right] \nonumber \\
&\times&
\frac{\left(\widehat{\dot{\Phi}}_1\right)^2} {\sqrt{1-k}}
+ \frac{\widehat{D_L}} {2(1+\widehat{z})} \frac{\widehat{\Phi}_1} {\sqrt{1-k}} \bigg[\frac{k'}{\widehat{\Phi}} - 
2 \frac{E'/E}{\widehat{\Phi}} (\widehat{\dot{\Phi}}^2 +k\nonumber \\
&-&   1)+ \widehat{\ddot{\Phi}'} + 2 \frac{\widehat{\dot{\Phi}'} \widehat{\dot{\Phi}}} 
{\widehat{\Phi}} - \widehat{\ddot{\Phi}} E'/E  \nonumber \\
&+&  \frac{1-k}{\left(\widehat{\Phi}_1\right)^2} \left( 3
\widehat{\ddot{\Phi}}E'/E -2 \frac{\widehat{\Phi'}\widehat{\ddot{\Phi}}} {\widehat{\Phi}} - \widehat{\ddot{\Phi}'} \right) \bigg] = 0.
\label{pparte}
\end{eqnarray}

To obtain a second order differential equation for $k$, we substitute into (\ref{pparte}) $E'/E$, as given by (\ref{E}), 
and $(E'/E)'$ and the partial derivatives of $\Phi$ as given in the Appendix. For readability purpose, we leave this task 
to the interested reader. These are functions of $k$, $k'$, $k''$, $M$, $\widehat{\rho}$, ${\rm d} \widehat{\rho}/{\rm d} z$, 
$\widehat{z}$ and the total derivatives of $\Phi$ with respect to $r$ up to second order.

To be able to solve for $M$ and $k$ we need a third differential equation involving $\widehat{\Phi}$ and possibly $M$ and $k$. 
It is given by (\ref{red}) into which we substitute the expressions for the partial derivatives of $\widehat{\Phi}$ as given in 
the Appendix and that of $E'/E$ as given by (\ref{E}). We obtain
\begin{eqnarray}
&& \frac{\sqrt{1-k}}{\left(1 + \widehat{z}\right)} \frac{{\rm d} \widehat{z}}{{\rm d} r} = \frac{4 M}{\widehat{\Phi}^2} + 
\frac{1}{\sqrt{\frac{2M}{\widehat{\Phi}} - k}} \left[\frac{2 M}{\widehat{\Phi}^2}\frac{{\rm d} \widehat{\Phi}}{{\rm d} r} \right. \nonumber \\
&+& \left. \sqrt{1-k}\left(4 \pi \widehat{\rho} \widehat{\Phi} - \frac{3 M}{\widehat{\Phi}^2}\right) - \frac{k'}{2}\right]  \nonumber \\
&-& \frac{1}{\widehat{\Phi}} \sqrt{\frac{2M}{\widehat{\Phi}} - k} \left[\frac{{\rm d} \widehat{\Phi}}{{\rm d} r}
 + \sqrt{\frac{2M}{\widehat{\Phi}} - k} - \sqrt{1-k}\right].
\label{dphi}
\end{eqnarray}

We have thus obtained a set of three differential equations for the unknown functions $M$, $k$ and $\widehat{\Phi}$. 
However, since the observables, $\widehat{\rho}$, ${\rm d} \widehat{\rho}/{\rm d} z$, $\widehat{D_L}$ and derivatives 
with respect to $z$ and $\widehat{z}$ are given in terms of the redshift $z$, and not of the unobservable coordinate $r$, 
we need to write them as functions of $r$ instead of $z$. This can be performed by using (\ref{red}) and the solution 
$\widehat{z}(r)$ of (\ref{ddrift}) given in below Sec.~\ref{zrdrift}. For readability purpose, we do not write down 
here the resulting equation and let the interested reader do the straightforward calculation for herself.

We solve numerically the set of three coupled equations (\ref{diffM}), (\ref{pparte}) and (\ref{dphi}) with the origin 
conditions $M(r=0) = k(r=0) = \widehat{\Phi}(r=0) = 0$ \cite{BKHC10} and obtain $M(r)$,  $k(r)$ and $\widehat{\Phi}(r)$. 
Then, $E(r)$ proceeds from (\ref{E}).

We have now at our disposal three arbitrary functions, $M(r)$, $k(r)$ and $E(r)$ which are sufficient for fully determining 
the model with our coordinate choice (\ref{coord}). The last integration function $t_B(r)$ proceeds from $M(r)$, $k(r)$ and 
$\widehat{\Phi}$ as follows.

We solve for $\widehat{\eta}$ from the first equation in (\ref{par})
and (\ref{para}) -- (\ref{parc}). Then $t_B$ follows from
\begin{equation}
\widehat{t} - t_B = t_0 - r - t_B = \frac{M}{K^{3/2}} \widehat{\xi}.
\label{tb}
\end{equation}

\subsection{Equation for $\widehat{z}(r)$} \label{zrdrift}

Besides $\widehat{\rho}$ and $\widehat{D_L}$, another measurable quantity we have at our disposal is the redshift-drift as 
given by (\ref{eq19}). On the light cone, after replacing the partial derivatives of $\widehat{\Phi}$ by their expressions 
given in the Appendix and using (\ref{red}), this equation can be written
\begin{equation}
\frac{{\rm d^2} \widehat{z}}{{\rm d} r^2} - \frac{1}{1 + \widehat{z}} \left(\frac{{\rm d} \widehat{z}}{{\rm d} r}\right)^2 + 
\left[ \left(1 + \widehat{z}\right) \frac{{\rm d} \widehat{\dot{z_0}}}{{\rm d} z} - \widehat{\dot{z_0}}\right]\frac{{\rm d} 
\widehat{z}}{{\rm d} r} = 0,
\label{ddrift}
\end{equation}
where $\widehat{\dot{z_0}}(z) = (\delta \widehat{z}/ \delta t_0)(z)$ is the redshift-drift for a 
given $\delta t_0$ as measured by the observer for a source at redshift $\widehat{z}$ on her past 
light cone when she is at $(t_0, r=0)$. This equation gives a second order differential equation 
for $\widehat{z}$. With the origin condition $\widehat{z}(r=0) = 0$ by definition of the redshift, 
it can be numerically solved from the measurements of $\widehat{\dot{z_0}}$ to yield $\widehat{z}(r)$. 
This allows us to write the $\widehat{\Phi}$ total derivatives with respect to $z$ instead of $r$,
 by using (\ref{red}).

\subsection{Algorithm}

To obtain the functions, $M$, $k$, $E$ and $t_B$ from the galaxy number count, $n(z)$, supernova 
luminosity distance and redshift-drift observations, we propose to proceed as follows:

\begin{enumerate}

\item take the discrete data $\ell(z, \theta, \phi)$ and $n(z, \theta, \phi)$, and correct them for 
known observational bias and selection effects. We do not consider here evolution effects. Average 
them over angles, fit them to some smooth analytic function of $z$ and obtain $\ell(z)$ and $n(z)$.

\item assuming a phenomenological $\widehat{L}(z)$ law, use (\ref{dl}) to obtain $\widehat{D_L}(z)$ 
and then its derivatives up to second order.

\item determine from the data a constant average galaxy mass $m$ (we do not consider here galaxy
 mass evolution) and obtain $\widehat{\rho}(z) = m n(z)$.

\item assuming observations of the redshift-drift have been performed over an elapsed time 
$\delta t_0$ sufficient to provide a robust data set, $\widehat{\dot{z}_0}(z, \theta, \phi)$, 
average it over all angles, fit it to some smooth analytic function of $z$ and obtain 
$\widehat{\dot{z}_0}(z)$ and its first derivative.

\item determine $\widehat{z}(r)$ by solving (\ref{ddrift}) and use it to transform the observables
 $\widehat{z}$, $\widehat{\rho}$, ${\rm d} \widehat{\rho}/{\rm d} z$, $\widehat{D_L}$ and its derivatives with respect to $z$ into functions of $r$.

\item determine $\widehat{\Phi}(r)$, $M(r)$ and $k(r)$ from the set of differential equations 
(\ref{dphi}), (\ref{diffM}) and (\ref{pparte}).

\item knowing $M(r)$, $k(r)$ and $\widehat{\Phi}(r)$, determine $E(r)$ by integrating (\ref{E}).

At this stage, we have obtained three arbitrary functions of $r$ which, 
with our coordinate choice, are enough to fully determine the model. However, for 
completeness, we give below the recipe to compute the last integration function $t_B(r)$.

\item solve for $\widehat{\eta}$ from the first equation in (\ref{par})
and (\ref{para}) - (\ref{parc}).

\item determine $t_B(r)$ from (\ref{tb}).

\end{enumerate}

This is the algorithm that could be applied to an ideal universe, where the redshift is monotonically 
increasing and where no shell-crossing is present to close the past null cone. We make such assumptions
 here but can be led to a more detailed study of the conditions imposed by them on the data in a future work.

Note that this recipe implies also the existence of a number of particular directions 
of axial symmetry in the observable Universe. It constitutes another simplifying assumption made at this stage.

To be complete, our above algorithm derivation should include a discussion of the 
existence and uniqueness of the differential equation set solutions. Given the difficulty of this task, 
we postpone it to future work and just assume here that solutions actually exist and that some of them, 
were they to be non unique, could be selected owing to their physical properties best designed to represent our Universe.

\subsection{Observables pertaining to the studied model}

\begin{figure}[!htb]
\begin{center}
 \includegraphics[width=9cm]{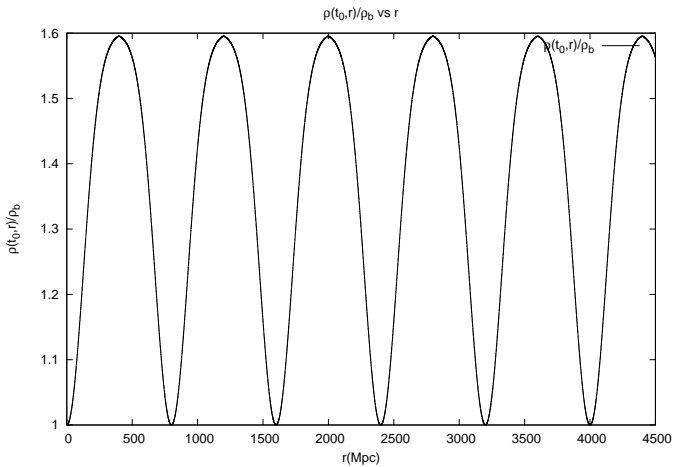}
\caption{The ratio of the QSS model mass density $\rho(t_0,r)$ and the background 
FLRW model mass density $\rho_b$ at the initial instant $t_0$ on the observer's world 
line. This quantity can be calculated but cannot be observed since it is in a 
spacelike relation to the observer.}
\label{rhovsr}
\end{center}
\end{figure}

\begin{figure}[!htb]
\begin{center}
 \includegraphics[width=9cm]{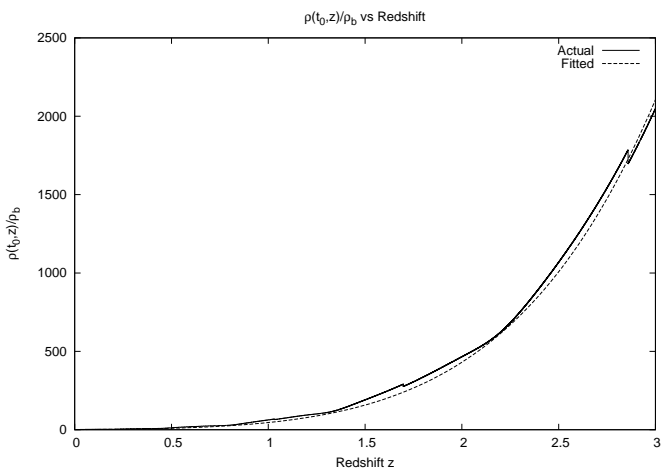}
\caption{The ratio of the QSS model mass density $\rho(t_0,z)$ and the background FLRW model 
mass density $\rho_b$ on the past light cone of the observer at O, before smoothing (solid curve) 
and after smoothing as described in the text (dotted curve).}
\label{rhovsz}
\end{center}
\end{figure}

The model studied in Sec.~\ref{sec4} has already been shown in BC to be able to 
reproduce to a good accuracy the supernova data, i.e., the observed luminosity distance-redshift relation. 
We have also depicted in Fig~\ref{drift_compfig} the redshift-drift for this model. 
To characterize it completely, we need thus the mass density $\rho$ on the past light 
cone of the observer at $O$, as shown in Sec.~\ref{sec5}. We have therefore computed it 
as a function of both the radial coordinate $r$ (Fig.~\ref{rhovsr}) and the redshift $z$ (Fig.~\ref{rhovsz}).

Notice that, while $\rho/\rho_b$ possesses a Swiss-cheese-like feature on the 
spacelike hypersurface $t=t_0$ (Fig.~\ref{rhovsr}), 
it becomes smoother on the past light cone, while exhibiting anyhow some cusps 
at the patch borders where the matching conditions are not perfect as already 
discussed in Sec.~\ref{sec4} (Fig.~\ref{rhovsz}).
 We have been able however to fit it rather accurately to the smooth 
$(1+z)^{5.5}$ curve (Fig.~\ref{rhovsz}) with maximum fractional error of $38\%$ and mean fractional error of $20\%$.

The redshift-drift appears smoother. Here, the non perfect matching is 
less visible. While all the model determining functions are constructed as 
in a Swiss-cheese, Fig.~\ref{drift_compfig} giving the drift on the past 
light cone from $O$ shows a very smooth curve. We have thus been able to 
fit it to the smooth $az^b+cz^d$ function with best fit for $a=-0.023, b=2.46, c=-0.029$ and $d=1.2$ with 
maximum fractional error of $12\%$ and mean fractional error of $2\%$.

Hence, it is interesting to note that a very patchy underlying 
inhomogeneous model can correspond to smooth quantities measured on 
the observer's past light cone. Such a result might be an interesting 
outcome of algorithms of the kind proposed here, were the Universe such 
that very inhomogeneous solutions able to represent it would correspond to smooth observables.

\section{Conclusions} \label{sec6}

The redshift-drift was recently proposed in the literature as a way to discriminate between 
different universe models able to reproduce the current cosmological data without resort to any 
dark energy component \cite{Uzan08,QA10,Yoo11}. It is considered as an interesting tool for 
getting rid of the degeneracy which can appear between different solutions. Its expression has been known for a long time 
in the Friedmannian framework and it was lately calculated for the inhomogeneous 
spherically symmetric LTB models  \cite{Yoo11}.

We have, in Sec.~\ref{sec3}, generalized this calculation to the axially symmetric QSS model and 
given there a new analytic expression for the drift in this model.

Then, we have applied this result to its numerical computation for the axially symmetric QSS 
Swiss-cheese BC model 5 which was shown in \cite{BC10} to be able to reproduce to a good accuracy 
the supernova data. Comparing it to the $\Lambda$CDM drift up to a redshift $z=3$, we have shown 
that this effect can be a good discriminator between both models since, (i) as it is well-known, 
the drift in the $\Lambda$CDM model is positive up to $z=2.5$, while it is negative with an 
enhanced decline at high $z$ in our model, (ii) at redshift $z=3$, where the drift has become 
negative for the $\Lambda$CDM model, its amplitude is much higher by a factor of $\sim 14$ in our model.

We have also compared the drift obtained for our model with that displayed in \cite{QA10,Yoo11} 
for LTB models. We have found that the redshift-drift in our model exhibits the same general shape as in these models, but that its magnitude is higher on average by a factor of 
about two at a given redshift. This has been discussed at length in Sec.~\ref{sec4}.

However, inhomogeneous exact models have been sometimes criticized on the ground that they exhibit 
more degrees of freedom than FLRW models and are thus able to fit more easily the data. Therefore, 
in Sec.~\ref{sec5} we have addressed the issue of finding how many and which independent data sets
 we need to be able to fully reconstruct an axially symmetric QSS model from background observations 
and we have also derived an algorithm for implementing this goal. Of course, we can presume that the 
choice of the data sets able to constrain the model
 is not unique, but the observables we have used are the only ones which have allowed us to derive 
a detailed algorithm for dealing so far with the issue.

Given the rather high amplitude of the drift in our model, some cosmologists might be tempted to 
use it carelessly as an actual discriminator between QSS Swiss-cheeses and other models. It was 
indeed studied in \cite{CAB07,VO07,LGV08} how the drift could be measured in the future by the 
proposed E-ELT instrument CODEX or in \cite{CAB07,VO07} by the VLT instrument ESPRESSO. These 
were claimed in \cite{QA10} to be able to discriminate between $\Lambda$CDM and different LTB 
models in the course of an observation decade. Therefore, since the drift in our model is higher 
by a factor of two, some researchers might be lead to conclude that a five year observation period 
might be sufficient to complete a test for our model. Another possibility to discriminate between 
inhomogeneous and homogeneous Universe models by measuring the redshift-drift with the future 
space-borne gravitational wave interferometer DECIGO/BBO was put forward in \cite{YNY11,YNY12} 
where the experiment duration was estimated to be around 5-10 years.

Now, we want to stress that the Swiss-cheese model we studied here is a mere toy model, not 
designed to be actually put to the test. We are not claiming that our Universe is constructed 
with axially symmetric patches, nor that we might be located at the origin of one of such 
patches or that the light emitted by distant sources should come to us following axially 
directed geodesics. 

However, we believe that cosmology is a very exciting science which is still in its infancy. 
For more than fifty years, homogeneous cosmology has been the main subject of study and, 
despite a number of attractive achievements, it is still facing huge unresolved problems, 
such as dark matter, dark energy, galaxy formation, etc. Inhomogeneous cosmology is still 
younger and it needs to progress step by step. This is the reason why, even if after LTB models, 
Szekeres ones are now coming slowly into play, their tricky properties lead us to begin with the 
study of the simpler QSS models, hence axial symmetry and its mathematical simplifications.

We are confident that the rather simple algorithm we display here to completely determine an 
axially symmetric QSS model from background observations will be generalized in the future to 
the most general QSS model, allowing therefore to obtain a very interesting representation of 
our Universe from cosmological data. We are determined to go on working in this direction.

\begin{acknowledgments}
One of us, MNC, wishes to thank Krzysztof Bolejko for helpful discussions about the properties 
of the BC model used in this paper and Andrzej Krasi\'nski for enlightening exchanges about QSS 
axially symmetric models. One of us, PM, wishes to thank Vincent Reverdy for a nice discussion 
about the numerical solution to the $k(r)$ determination. We would like to thank also Miguel Quartin and 
Chul-Moon Yoo for sending us the data files of Fig. 7 in \cite{QA10} and Fig. 5(b) in \cite{Yoo11}, respectively.

\end{acknowledgments}

\section*{Appendix}
\appendix

We give here the expressions of the partial derivatives of $\widehat{\Phi}$ needed to perform the calculations in Sec.~\ref{sec5}.

The first derivative with respect to time is already known from (\ref{dotphi2}). We recall it here for completeness:
\begin{equation}
\widehat{\dot{\Phi}} = \sqrt{\frac{2M}{\widehat{\Phi}} - k}.
\label{dotphi3}
\end{equation}

To obtain $\ddot{\Phi}$ on the light cone, we take the derivative of (\ref{hatdot}) 
with respect to the time coordinate which gives
\begin{equation}
\widehat{\ddot{\Phi}} = - \frac{M}{\widehat{\Phi}^2}.
\label{phidd}
\end{equation}

From (\ref{totder}) with our coordinate choice (\ref{coord}) and (\ref{dotphi3}) for $\widehat{\dot{\Phi}}$, it comes
\begin{equation}
\widehat{\Phi'} =  \frac{{\rm d} \widehat{\Phi}}{{\rm d} r} + \widehat{\dot{\Phi}} .
\label{phipr}
\end{equation}

To obtain $\widehat{\dot{\Phi}'}$, we differentiate (\ref{dotphi3}) with respect to $r$, 
then substitute $\widehat{\Phi'}$ as given by (\ref{phipr}) and $M'$ as given by (\ref{diffM}) 
into the resulting equation, which gives
\begin{equation}
 \widehat{\dot{\Phi}'} = \frac{2 M}{\widehat{\Phi}^2} + \frac{1}{\sqrt{\alpha} }
 \left[\frac{2 M}{\widehat{\Phi}^2}\frac{{\rm d} \widehat{\Phi}}{{\rm d} r} \right. 
+ \left. \sqrt{1-k}\beta - \frac{k'}{2}\right],
\label{phidpr}
\end{equation}

where
\begin{equation}
\alpha \equiv \frac{2 M}{\widehat{\Phi}} -k
\end{equation}

and 
\begin{equation}
 \beta \equiv 4 \pi \widehat{\rho} \widehat{\Phi} - \frac{3M}{\widehat{\Phi}^2}.
\end{equation}
Now, we wish to calculate $\widehat{\Phi''}$. With our coordinate choice (\ref{coord}), we can write
\begin{equation}
\widehat{\Phi'} = \sqrt{1-k} + \widehat{\Phi}E'/E,
\label{phipprint}
\end{equation}
which we differentiate with respect to $r$ to obtain
\begin{equation}
\widehat{\Phi''} = - \frac{k'}{2\sqrt{1-k}} + \widehat{\Phi'}E'/E + \widehat{\Phi}(E'/E)'.
\label{phipprint2}
\end{equation}
The expression for $\widehat{\Phi''}$ is therefore obtained by substituting into (\ref{phipprint2}), 
the expression for $\widehat{\Phi'}$ as given by (\ref{phipr}) and those for $E'/E$  as 
given by (\ref{E}) and for $(E'/E)'$ as calculated below.

Differentiating (\ref{E}) with respect to $r$ and rearranging yields
\begin{eqnarray}
&&\left(\frac{E'}{E}\right)' = \frac{1}{\widehat{\Phi}} \frac{{\rm d^2} \widehat{\Phi}}{{\rm d} r^2} 
- \frac{1}{\widehat{\Phi}^2} \frac{{\rm d} \widehat{\Phi}}{{\rm d} r} \left( \frac{{\rm d} \widehat{\Phi}}{{\rm d} r}
 + \sqrt{\alpha}  - \sqrt{1 - k}\right) \nonumber \\
&+& \frac{1}{\sqrt{\alpha}} \left[\frac{2M}{\widehat{\Phi}^2} \frac{{\rm d} \widehat{\Phi}}{{\rm d} r} + \beta \sqrt{1 - k} \right]
+ \frac{k'}{2} \left(\frac{1}{\widehat{\Phi}\sqrt{1 - k}} - \frac{1}{\sqrt{\alpha}}\right)\nonumber \\
& +& \frac{3M}{\widehat{\Phi}^2}.
\label{Epr}
\end{eqnarray}

After some calculations, we thus obtain for $\widehat{\Phi''}$
\begin{eqnarray}
 \widehat{\Phi''} &=& \frac{{\rm d^2} \widehat{\Phi}}{{\rm d} r^2} + \frac{1}{\widehat{\Phi}} \left[3M + \alpha 
+ \sqrt{\alpha} \frac{{\rm d} \widehat{\Phi}}{{\rm d} r} - \sqrt{\alpha (1-k)}  \right] \nonumber \\
&+&\frac{1}{\sqrt{\alpha} } \left[\frac{2M}{\widehat{\Phi}} \frac{{\rm d} \widehat{\Phi}}{{\rm d} r} \right. 
+ \left. \sqrt{1 - k} \widehat{\Phi}\beta -\frac{k'}{2} \widehat{\Phi} \right].
\label{phippr}
\end{eqnarray}

Differentiating (\ref{phidd}) with respect to $r$ and substituting into the resulting equation 
the expressions for $\widehat{\Phi'}$ and $M'$ as given respectively by (\ref{phipr}) and (\ref{diffM}), we obtain
\begin{equation}
 \widehat{\ddot{\Phi}'} = - \frac{M}{\widehat{\Phi}^3} \left( \frac{{\rm d} \widehat{\Phi}}{{\rm d} r} + 
\sqrt{\alpha} \right) - \sqrt{1-k} \frac{\beta}{\widehat{\Phi}}.
\label{phiddpr}
\end{equation}

Finally, to obtain $\widehat{\dot{\Phi}''}$, we differentiate $\widehat{\dot{\Phi}'}$ with respect to $r$ and obtain
\begin{eqnarray}
&& \widehat{\dot{\Phi}''} = \frac{2M}{\widehat{\Phi}^2 \sqrt{\alpha} } \frac{{\rm d^2} \widehat{\Phi}}{{\rm d} r^2} 
+ \frac{2M}{\widehat{\Phi}^3} \left(3 \frac{{\rm d} \widehat{\Phi}}{{\rm d} r} + \sqrt{\alpha}  \right) \nonumber \\
&+& \sqrt{1-k} \left( 12 \pi \widehat{\rho} - \frac{11M}{\widehat{\Phi}^3}\right) - 
\frac{1}{\alpha^{3/2}} \left[\frac{2M}{\widehat{\Phi}^2} \left(\frac{{\rm d} \widehat{\Phi}}{{\rm d} r}+ \sqrt{\alpha} \right) \right. \nonumber \\
&+& \left.  \sqrt{1-k} \beta - \frac{k'}{2}\right]
 \left[\frac{2M}{\widehat{\Phi}^2} \frac{{\rm d} \widehat{\Phi}}{{\rm d} r} 
+ \sqrt{1-k} \beta - \frac{k'}{2}\right] \nonumber \\
&+& \frac{1}{\sqrt{\alpha}} \left[ \frac{2M}{\widehat{\Phi}^3}
\left(\frac{{\rm d} \widehat{\Phi}}{{\rm d} r}\right)^2 + \sqrt{1-k} \left(\frac{3 \beta}{\widehat{\Phi}}\frac{{\rm d} \widehat{\Phi}}{{\rm d} r} 
+  4 \pi \widehat{\rho'} \widehat{\Phi} \right) \right. \nonumber \\
&-&   3 (1-k) \frac{\beta}{\widehat{\Phi}} 
+ \left. \frac{2M}{\widehat{\Phi}^3}(3M -k) - \frac{k'}{2 \sqrt{1-k}} \beta - \frac{k''}{2} \right] \nonumber \\
&+& \frac{2M\left(\widehat{\Phi} - 1 \right)}{\alpha \widehat{\Phi}^2 }
 \left[ \frac{2M}{\widehat{\Phi}^2} \frac{{\rm d} \widehat{\Phi}}{{\rm d} r} + \sqrt{1-k}
\beta -\frac{k'}{2} \right],
\label{phidppr}
\end{eqnarray}
where we have replaced ${\partial}/{\partial r}({\rm d}\widehat{\Phi}/ {\rm d} r)$ by 
its expression $\widehat{\Phi''} - \widehat{\dot{\Phi}'}$ following from our coordinate 
choice, with $\widehat{\Phi''}$ and $\widehat{\dot{\Phi}'}$ as obtained from, respectively,
 (\ref{phippr}) and (\ref{phidpr}) substituted, and where we have also inserted $M'$ as given 
by (\ref{diffM}) and $\widehat{\Phi}'$ as given by (\ref{phipr}).

We see that this equation includes $\widehat{\rho'}$. We use (\ref{red}) and the expressions 
(\ref{phidpr}) of $\widehat{\dot{\Phi}'}$, and (\ref{phipr}) of $\widehat{\Phi'}$, to write 
$\widehat{\rho'}$ in terms of ${\rm d} \widehat{\rho} / {\rm d} z$ as
\begin{eqnarray}
&& \widehat{\rho'} = \frac{1 + \widehat{z}}{\sqrt{1-k}} \left\{ \frac{1}{\sqrt{\alpha}} 
\left[\frac{2M}{\widehat{\Phi}^2} \frac{{\rm d} \widehat{\Phi}}{{\rm d} r}  
+\sqrt{1-k} \beta - \frac{k'}{2}\right] \right. \nonumber \\
&-& \left. \frac{\sqrt{\alpha}}{\widehat{\Phi}} \left(\frac{{\rm d} \widehat{\Phi}}{{\rm d} r} 
- \sqrt{1-k}\right) + \frac{k}{ \widehat{\Phi}} \right\} \frac{{\rm d} \widehat{\rho}}{{\rm d} z}.
\label{rhopr}
\end{eqnarray}

Examining the above expressions for $(E'/E)'$ and the different partial derivatives of
 $\widehat{\Phi}$ involved in the set of three coupled differential equations derived 
in Sec.~\ref{sec5}, it is easy to conclude that these equations relate $\widehat{\Phi}$ 
and its total derivative with respect to $r$ up to second order, the observables 
$\widehat{\rho}(z)$ and its first derivative with respect to $z$, $\widehat{D_L}$ 
and its derivatives with respect to $z$ up to second order, and $\widehat{z}$, and 
the arbitrary functions with their derivatives, $M$, $M'$, $k$, $k'$, $k''$, which 
allows us to solve for $\widehat{\Phi}(r)$, $M(r)$ and $k(r)$ with the algorithm describe in Sec.~\ref{sec5}.


\end{document}